\documentclass[pra,aps,twocolumn,showpacs]{revtex4}
\usepackage{times,epsfig,amssymb,amsfonts,amsmath}
\newcommand{\ket}[1]{|#1\rangle}
\newcommand{\bra}[1]{\langle #1|}
\newcommand{\proj}[1]{\ket{#1}\bra{#1}}

\begin{document}

\title{Exploring multipartite quantum correlations with the square of quantum discord}

\author{Yan-Kui Bai$^{1,2}$}
\email{ykbai@semi.ac.cn}
\author{Na Zhang$^1$}
\author{Ming-Yong Ye$^{3,2}$}
\author{Z. D. Wang$^2$}
\email{zwang@hku.hk}

\affiliation{$^1$ College of Physical Science and Information
Engineering and Hebei Advance Thin Films Laboratory, Hebei Normal
University, Shijiazhuang, Hebei 050016, China\\
$^2$ Department of Physics and Center of Theoretical and Computational
Physics, The University of Hong Kong, Pokfulam Road, Hong Kong, China\\
$^3$ College of Physics and Energy, Fujian Normal
University, Fuzhou 350007, China}

\begin{abstract}
We explore the quantum correlation distribution in multipartite quantum states based
on the square of quantum discord (SQD). For tripartite quantum systems, we derive
the necessary and sufficient condition for the SQD to satisfy the monogamy
relation. Particularly, we prove that the SQD is monogamous for
three-qubit pure states, based on which a genuine tripartite quantum correlation
measure is introduced. In addition, we also address the quantum correlation
distributions in four-qubit pure states.  As an example, we investigate multipartite
quantum correlations in the dynamical evolution of multipartite cavity-reservoir
systems.
\end{abstract}


\maketitle

\section{introduction}

Besides quantum entanglement, quantum correlation is also a key resource in quantum
information processing
\cite{knill98prl,datta08prl,lanyon08prl,piani08prl,piani10prl,madhok11pra,
cav11pra,roa11prl,dakic12np,cui12natc,datta13ijmpb}. As a basic tool to characterize the quantum advantage
\cite{modi12rmp}, quantum discord (QD) is a prominent bipartite quantum correlation
measure \cite{ollivier01prl,vedral01jpa}. Recently, generalization of the QD to
multipartite systems has received much attention
\cite{modi10prl,rulli11pra,pati10arxiv,giorgi11prl,walczak11epl}. However,
characterization of quantum correlation structure in multipartite systems is still
very challenging. Monogamy relation \cite{ckw00pra,osb06prl,horodecki09rmp} is an
important property in multipartite quantum systems. As quantified by the square of
concurrences \cite{wootters98prl}, entanglement is monogamous in multiqubit systems
\cite{osb06prl} \emph{i.e.},
\begin{equation}\label{1}
C^2_{A_1|A_2\cdots A_N}\geq C^2_{A_1A_2}+C^2_{A_1A_3}+\cdots +C^2_{A_1A_N},
\end{equation}
and this property can be used to construct genuine multipartite entanglement
measures \cite{ckw00pra,byw07pra}. Therefore, it is natural to ask whether or not
the quantum correlation is monogamous, especially for the QD.

Prabhu \emph{et al.} found that the QD is not monogamous and the monogamy relation
\begin{equation}\label{2}
D_{A|BC}-D_{A|B}-D_{A|C}\geq 0
\end{equation}
is not satisfied even for the three-qubit $W$ state \cite{prabhu12pra}. Giorgi
\cite{giorgi11pra} and Fanchini \emph{et al} \cite{fanchini11pra,fanchi11arxiv}
related the monogamy condition of QD to the entanglement of formation, while Ren and Fan
showed that QD is not monogamous under the same measurement party \cite{ren11arxiv}.
Recently, Streltsov \emph{et al} further showed that the monogamy relation does not
hold in general for quantum correlation measures which are nonzero for separable
states \cite{str12prl}. However, these results do not imply that quantum correlation
is still not monogamous in a specific case (for example, the geometric measure of
discord \cite{dakic10prl} is monogamous in three-qubit pure states \cite{str12prl}).
Since the QD is accepted as a basic tool for quantum correlation, it is desirable to
find a kind of monogamous QD even in several qubit systems, which on the one hand
gives a clear correlation structure but on the other hand allows the
characterization of genuine multipartite quantum correlation.

In this paper, we are motivated by the following two questions: (i) \emph{whether or
not} the QD is monogamous in certain form, and (ii) \emph{in what degree} the discord is
monogamous and can characterize the genuine multipartite quantum correlation. To
answer these two questions, we explore the monogamy property of the square of
quantum discord (SQD) in multipartite quantum systems. The paper is organized as
follows. In Sec. II, we derive the necessary and sufficient condition for the
SQD to be monogamous in tripartite quantum states. In three-qubit pure states, we prove
that the SQD is monogamous and define a genuine tripartite quantum correlation
measure. In Sec. III, we analyze the correlation distribution in multiqubit pure
states and construct multipartite quantum correlation indicators. As an application,
we address the dynamics of quantum correlation in multipartite cavity-reservoir
systems. Finally, we present discussions and a conclusion in Sec. IV.

\section{Monogamy property and correlation measure in tripartite quantum states}

\subsection{Definitions and monogamous condition}

In a bipartite quantum system $\rho_{AB}$, the total correlation can be quantified by
quantum mutual information $I_{A:B}=S(A)+S(B)-S({AB})$ with
$S(X)=-\mbox{Tr}\rho_{X}\mbox{log}_2\rho_{X}$ being von Neumann entropy
\cite{ollivier01prl}, while the classical correlation is given by
$J_{A:B}=\mbox{max}_{\{E_{j}^{B}\}}[S(A)-\sum_j p_j S(A|E_{j}^{B})]$, in which
$\{E_{j}^{B}\}$ is a positive operator-valued measure (POVM) performed on the
subsystem $B$ and
$\rho_{A|E_{j}^{B}}=\mbox{Tr}_{B}(E_{j}^{B}\rho_{AB}E_{j}^{B\dagger})/p_j$ with
$p_j=\mbox{Tr}_{AB}(E_{j}^{B}\rho_{AB}E_{j}^{B\dagger})$ \cite{vedral01jpa}. The QD
is used to characterize bipartite quantum correlation, which is defined as the
difference between $I_{A:B}$ and $J_{A:B}$, and is expressed as \cite{ollivier01prl}
\begin{equation}\label{3}
  D_{A|B}=S(B)-S(AB)+\mbox{min}_{\{E_{j}^{B}\}}\sum_j p_j
  S(A|E_{j}^{B}),
\end{equation}
where the minimum runs over all the POVMs, and $D_{A|B}$ is referred to as the
discord of system $AB$ with the measurement on subsystem $B$. The QD can also be
written in the form of quantum conditional entropy \cite{cav11pra}:
\begin{equation}\label{4}
  D_{A|B}=\widetilde{S}(A|B)-S(A|B),
\end{equation}
where the non-negative quantity $\widetilde{S}(A|B)=\mbox{min}_{\{E_{j}^{B}\}}\sum_j
p_j S(A|E_{j}^{B})$ is the measurement-induced quantum conditional entropy and
$S(A|B)=S(AB)-S(B)$ is the direct quantum generalization of conditional entropy.

Monogamy relation is an important property in multipartite quantum systems. Coffman
\emph{et al.} first showed that the monogamy relation of concurrence
$\mathcal{C}^2_{A|BC}-\mathcal{C}^2_{AB}-\mathcal{C}^2_{AC}\geq 0$
is satisfied in three-qubit quantum states and the residual entanglement can
characterize the genuine tripartite entanglement \cite{ckw00pra}. It should be
noted that, in the monogamy relation, the square of concurrence is monogamous other
than the concurrence itself which is not monogamous.
Previous studies indicated that the QD is not monogamous even in three-qubit pure
states \cite{prabhu12pra,giorgi11pra,fanchini11pra,fanchi11arxiv,ren11arxiv},
which does not imply that the square of QD is not monogamous either.

Here, we explore the monogamy property of SQD in multipartite systems. The SQD can be
written as
\begin{equation}\label{5}
D_{A|B} ^2=[\widetilde{S}(A|B)-S(A|B)]^2,
\end{equation}
which satisfies all the standard requirements for quantum correlation measure
\cite{str12prl,brodutch12qic} and can characterize effectively quantum correlation
in bipartite systems. Particularly, in a tripartite pure state $\ket{\psi_{ABC}}$,
the measurement-induced quantum conditional entropies are related to the entanglement
of formation \cite{wootters98prl} by the Koashi-Winter formula \cite{koashi04pra}
\begin{equation}\label{6}
  \widetilde{S}(i|k)=\widetilde{S}(j|k)=E_{f}(ij),
\end{equation}
where $\widetilde{S}(i|k)$ and $\widetilde{S}(j|k)$ are the conditional entropies
with measurement on the subsystem $k$, and $E_{f}(ij)=\mbox{min}\sum_{\epsilon}
p_\epsilon S(\rho_i^\epsilon)$ is the entanglement of formation in the subsystem
$\rho_{ij}$ with the minimum taking over all the pure state decompositions
$\{p_\epsilon, \rho_{ij}^\epsilon\}$ and $i\neq j\neq k \in \{A,B,C\}$. Using the
formula in Eq. (6), the SQD has the form
\begin{equation}\label{7}
  D_{i|k}^2=[E_f(ij)-S(i|k)]^2,
\end{equation}
where the measurement is performed on subsystem $k$, and $i\neq j\neq
k\in\{A,B,C\}$. Moreover, in a tripartite pure state $\ket{\psi_{ABC}}$, we have
the relation $D_{A|BC}^2=S^2(A)=E_f^2(A|BC)$ in which $E_f(A|BC)$ is the entanglement
of formation under the bipartite partition $A|BC$ \cite{ollivier01prl,vedral01jpa}.
Combining this relation with Eq. (7), we can derive
the quantum correlation distribution of SQD
\begin{eqnarray}\label{8}
 &&D_{A|BC}^2-D_{A|B}^2-D_{A|C}^2= T_{1}+T_{2},
\end{eqnarray}
where
\begin{eqnarray}\label{9}
 T_{1}&=& E_f^2(A|BC)-E_f^2(AB)-E_f^2(AC),\nonumber\\
 T_{2}&=& 2S(A|B)[E_f(AC)-E_f(AB)-S(A|B)].
\end{eqnarray}
In the distribution, the first term $T_{1}$ is an entanglement distribution relation
quantified by the square of entanglement of formation $E_f^2$ and the second term
$T_{2}$ is a function of entanglement of formation $E_f$ and conditional entropy
$S(A|B)$. According to Eq. (8), the necessary and sufficient condition for the monogamous SQD is
\begin{equation}\label{10}
  T_{1}+T_{2} \geq 0.
\end{equation}

\subsection{Monogamy property in three-qubit pure states}

We now look into the quantum correlation distribution in two-level (qubit) systems.

\emph{Theorem I.}  In any three-qubit pure state $\ket{\psi_{ABC}}$, the square of
quantum discord $D_{A|BC}$ obeys the monogamy relation
\begin{equation}\label{11}
    D^2_{A|BC}-D^2_{A|B}-D^2_{A|C}\geq 0.
\end{equation}

\emph{Proof.} The theorem will hold when the monogamy condition in Eq. (10) is
satisfied for all three-qubit pure states. In two-qubit quantum states, the
entanglement of formation has an analytical expression
$E_f(\rho_{ij})=h[(1+(1-C_{ij}^2)^{1/2})/2]$ in which
$h(x)=-x\mbox{log}_2x-(1-x)\mbox{log}_2(1-x)$ is the binary entropy and
$C_{ij}=\mbox{max}\{0,\sqrt{\lambda_1}-\sqrt{\lambda_2}-\sqrt{\lambda_3}-\sqrt{\lambda_4}\}$
is the concurrence with the decreasing non-negative $\lambda_i$s being the
eigenvalues of matrix $\rho_{ij}(\sigma_y\otimes
\sigma_y)\rho_{ij}^*(\sigma_y\otimes \sigma_y)$ \cite{wootters98prl}. As a function
of the square of concurrence, the entanglement of formation obeys the following
relations:
\begin{eqnarray}\label{12}
  E_f^2(C^2_{A|BC})&\geq& E_f^2(C^2_{AB}+C^2_{AC})\nonumber\\
  &\geq& E_f^2(C^2_{AB})+E_f^2(C^2_{AC}),
\end{eqnarray}
where the Coffman-Kundu-Wootters (CKW) relation $C^2_{A|BC}\geq C^2_{AB}+C^2_{AC}$ \cite{ckw00pra} and the
monotonically increasing property of $E_f(C^2)$ is used in the first equation, and
the property that $E_f^2$ is a convex function of $C^2$ is used in the second
equation. According to Eq. (12), we can obtain the first term $T_{1}\geq 0$ in the
monogamy condition.

For the second term $T_{2}$, we first show that  $[E_f(AC)-E_f(AB)]$ has the same
sign as that of $S(A|B)$. It is straightforward to derive the following relations:
\begin{eqnarray}\label{13}
  E_f(C^2_{AC})\geq E_f(C^2_{AB})
  &\Rightarrow& E_f(C^2_{AB|C})\geq E_f(C^2_{AC|B})\nonumber\\
  &\Rightarrow& S(C)\geq S(B)\nonumber\\
  &\Rightarrow& S(A|B)\geq 0,
\end{eqnarray}
where we have used the entanglement distributions
$C^2_{AB|C}=C^2_{AC}+C^2_{BC}+\tau_3$ and $C^2_{AC|B}=C^2_{AB}+C^2_{BC}+\tau_3$ with
$\tau_3$ being the three-tangle \cite{ckw00pra}, and the monotonically increasing
property of $E_f(C^2)$. Similarly, if $E_f(AC)-E_f(AB)\leq 0$, we can obtain the
relation $S(A|B)\leq 0$. Therefore $[E_f(AC)-E_f(AB)]$ and $S(A|B)$ have the same
sign, and thus the second term in the monogamy condition has the form
\begin{eqnarray}\label{14}
 T_{2}=2|S(A|B)|[|E_f(AC)-E_f(AB)|-|S(A|B)|].
\end{eqnarray}
As a result, the non-negative property of $T_{2}$ is equivalent to
\begin{equation}\label{15}
 T_{2}^{\prime}=|E_f(AC)-E_f(AB)|-|S(A|B)|\geq 0,
\end{equation}
which is proven to be valid as follows.

On one hand, if $E_f(AC) \geq E_f(AB)$, the left-hand side of Eq.(15) can be written as
\begin{eqnarray}\label{16}
  T_{2}^{\prime}(+)=S(B)-E_f(AB)-S(C)+E_f(AC),
\end{eqnarray}
where we have used  $S(A|B)=S(C)-S(B)$ in tripartite pure states. On the other hand,
we have
\begin{eqnarray}\label{17}
&&E_f(C_{AC}^2)\geq E_f(C_{AB}^2) \nonumber\\
&\Rightarrow& E_f(C_{AC}^2+\Delta)\geq  E_f(C_{AB}^2+\Delta)\nonumber\\
&\Rightarrow& E_f(C_{AC}^2+\Delta)-E_f(C_{AC}^2)\nonumber\\
&&\leq
E_f(C_{AB}^2+\Delta)-E_f(C_{AB}^2),
\end{eqnarray}
where $\Delta$ is a non-negative constant. In addition, we have used the monotonic
property of $E_f(C^2)$ in the second inequality and the concave property of
$E_f(C^2)$ \cite{giorgi11pra} in the third inequality which means that along with
the increase of concurrence $C^2$ the increment of $E_f$ will decrease. When we
choose $\Delta=C_{BC}^2+\tau_3$, the entanglement of formation is
\begin{eqnarray}\label{18}
 E_f(C_{AC}^2+\Delta)&=&E_f(C_{AC}^2+C_{BC}^2+\tau_3)\nonumber\\
 &=&E_f(C_{C|AB}^2)\nonumber\\
 &=&S(C),
\end{eqnarray}
where the CKW relation has been used. Similarly, the relation
$E_f(C_{AB}^2+\Delta)=S(B)$ can be derived. Substituting the results into Eq. (17),
we have the relation
\begin{equation}\label{19}
S(B)-E_f(AB)\geq S(C)-E_f(AC).
\end{equation}
Combining Eq. (19) with Eq. (16), we can obtain $T_{2}^{\prime}(+)\geq 0$. In the
other case, if $E_f(AC)\leq E_f(AB)$, the left-hand side of Eq. (15) becomes
\begin{equation}\label{20}
    T_{2}^{\prime}(-)=S(C)-E_f(AC)-S(B)+E_f(AB).
\end{equation}
Moreover, we have
\begin{eqnarray}\label{21}
&&E_f(C_{AC}^2)\leq E_f(C_{AB}^2) \nonumber\\
&\Rightarrow& E_f(C_{AC}^2+\Delta)-E_f(C_{AC}^2)\nonumber\\
&&\geq E_f(C_{AB}^2+\Delta)-E_f(C_{AB}^2)\nonumber\\
&\Rightarrow& S(C)-E_f(AC)\geq S(B)-E_f(AB),
\end{eqnarray}
where  $\Delta=C^2_{BC}+\tau_3$ and $E_f(C_{Ak}^2+\Delta)=S(k)$ with $k\in\{B,C\}$, and
the concave property of $E_f(C^2)$ is used. Combining Eq. (20)
with Eq. (21), we get $T_{2}^{\prime}(-)\geq 0$. Therefore, we have proven that
$T_{2}^{\prime}$ is non-negative, namely, $T_{2}$ is non-negative. Due to $T_{1}\geq
0$ and $T_{2}\geq 0$, the monogamy condition holds, and the proof is completed.

\begin{figure}
\begin{center}
\epsfig{figure=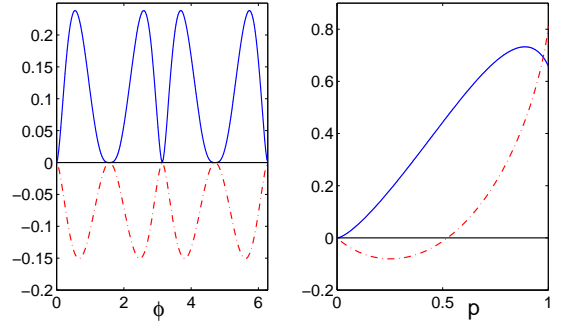,width=0.4\textwidth}
\end{center}
\caption{(Color online) Quantum correlation distribution of SQD (blue solid
line) in comparison to that of QD (red dash-dotted line). Left: two distributions
for the generalized $W$ state in Eq. (22) as a function of parameter $\phi$ where the
parameter $\theta$ is set to $\pi/4$; Right: two distributions for the two-parameter
state in Eq. (23) as a function of the parameter $p$ where the other parameter
is chosen to be $\epsilon=0.5$.}
\end{figure}

As examples, we consider the quantum correlation distribution of SQD in generalized
$W$ state \cite{prabhu12pra}
\begin{equation}\label{22}
\ket{\psi_W}=\mbox{sin}\theta\mbox{cos}\phi\ket{011}+\mbox{sin}\theta\mbox{sin}\phi
\ket{101}+\mbox{cos}\theta\ket{110}
\end{equation}
and the two-parameter state \cite{giorgi11pra}
\begin{eqnarray}\label{23}
\ket{\psi(p,\epsilon)}&=&\sqrt{p\epsilon}\ket{000}
+\sqrt{p(1-\epsilon)}\ket{111}\nonumber\\
&&+\sqrt{(1-p)/2}(\ket{101}+\ket{110}).
\end{eqnarray}
In Fig.1, we plot the distribution
$D^2_{A|BC}-D^2_{A|B}-D^2_{A|C}$ (blue solid line) in comparison to the distribution
$D_{A|BC}-D_{A|B}-D_{A|C}$ (red dash-dotted line) for the two quantum states, where
although the QD is not monogamous as pointed out in Refs.
\cite{prabhu12pra,giorgi11pra}, we can see that the SQD is monogamous.

For the further verification on the theorem, we analyze the standard form of three-qubit
pure states \cite{acin00prl}
\begin{eqnarray}\label{24}
  \ket{\Psi}_{ABC}&=&\lambda_0
  \ket{000}+\lambda_1e^{i\phi}\ket{100}+\lambda_2\ket{101}+\lambda_3\ket{110}\nonumber\\
  &&+\lambda_4\ket{111},
\end{eqnarray}
where the real number $\lambda_i$ ranges in $[0,1]$ with the condition $\sum
\lambda_i^2=1$, and the relative phase $\phi$ changes in $[0,\pi]$. Without loss of
generality, we set $\lambda_0=\mbox{cos}\theta_0$,
$\lambda_1=\mbox{sin}\theta_0\mbox{cos}\theta_1$,
$\lambda_2=\mbox{sin}\theta_0\mbox{sin}\theta_1\mbox{cos}\theta_2$,
$\lambda_3=\mbox{sin}\theta_0\mbox{sin}\theta_1\mbox{sin}\theta_2\mbox{cos}\theta_3$,
and $\lambda_4=\mbox{sin}\theta_0
\mbox{sin}\theta_1\mbox{sin}\theta_2\mbox{sin}\theta_3$, respectively. In Fig.2, the
quantum correlation distribution of SQD is plotted as a function of parameters
$\theta_0, \theta_1,\theta_2$, and $\theta_3$ (the relative phase is set to
$\phi=0$), where $\theta_i$ ranges in $[0,\pi/2]$ with equal interval being $\pi/40$.
Again, we can see that the SQD is monogamous.

\begin{figure}
\begin{center}
\epsfig{figure=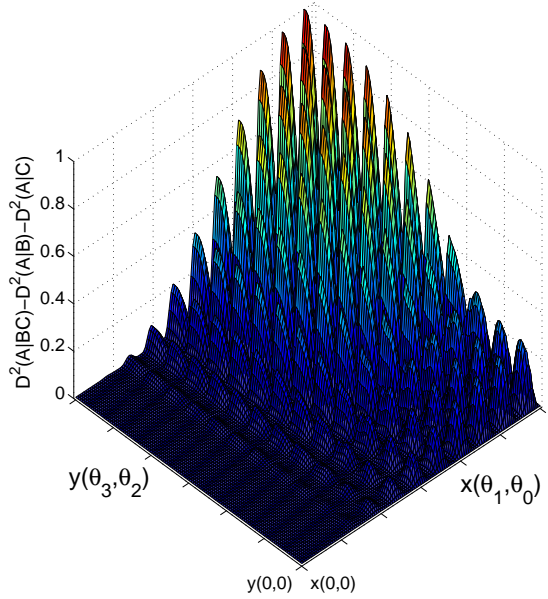,width=0.4\textwidth}
\end{center}
\caption{(Color online) The monogamy property of SQD for the standard form of
three-qubit pure states in Eq. (24). The distribution of SQD is plotted as a
function of $x(\theta_1, \theta_0)$ and $y(\theta_3,\theta_2)$ where $\theta_i$
ranges in $[0,\pi/2]$ with equal interval being $\pi/40$ and the relative phase is
set to $\phi=0$.}
\end{figure}

\subsection{A genuine three-qubit quantum correlation measure
with the hierarchy structure}

A quantum correlation measure should satisfy the following necessary criteria: (i)
it should be a non-negative real number; (ii) it is invariant under local unitary
operations \cite{str12prl,brodutch12qic}; and (iii) it is zero in an $n$-partite
quantum state if and only if the state is a product state in any bipartite cut
\cite{ben11pra}.

Based on our previous analysis on the quantum correlation distribution of SQD, we
define a tripartite quantum correlation measure as
\begin{equation}\label{25}
  Q_3(A|BC)=D^2_{A|BC}-D^2_{A|B}-D^2_{A|C},
\end{equation}
which characterizes the genuine three-qubit quantum correlation in a pure state
$\ket{\psi_{ABC}}$. The non-negative property of $Q_3$ is satisfied due to the SQD
being monogamous. The tripartite correlation $Q_3$ is invariant under local unitary
operations because the SQDs are unchanged under the transformation.

For the third requirement, we first prove that the measure $Q_3(A|BC)$ is zero if a
three-qubit state is a product state in any bipartite cut. When the quantum state
has the form $\ket{\psi_{ABC}}=\ket{\varphi_A}\otimes\ket{\varphi_{BC}}$, the SQD
$D^2_{A|BC}=S^2(A)=0$ due to the product property under this partition.
The SQD $D^2_{A|B}=0$ because we have $\sum(I_A\otimes E^B_j)\rho_{AB}(I_A\otimes
E^{B\dagger}_j)=\rho_{AB}$ with $E^B_j$ being the projector composed of the
eigenvector of $\rho_B$. The case for $D^2_{A|C}=0$ is similar. So, the genuine
tripartite quantum correlation $Q_3(A|BC)=0$. For the product state
$\ket{\psi_{ABC}^{\prime}}=\ket{\varphi_{AB}}\otimes\ket{\varphi_{C}}$, we also have
$Q_3(A|BC)=0$, since $D^2_{A|BC}=D^2_{A|B}=S^2(A)$ and $D^2_{A|C}=0$. Similarly, we
can derive $Q_3(A|BC)=0$ for
$\ket{\psi_{ABC}^{\prime\prime}}=\ket{\varphi_{AC}}\otimes\ket{\varphi_{B}}$.
Therefore, $Q_3(A|BC)$ is zero when the three-qubit pure state is a product state in any
bipartite cut.

Next, we prove that when the three-qubit pure state is not a bipartite product under
any partition, the measure $Q_3$ is always nonzero. Based on the correlation
distribution in Eq. (8), it is sufficient to prove the term
$T_1=E_f^2(C_{A|BC}^2)-E_f^2(C_{AB}^2)-E_f^2(C^2_{AC})>0$ since the second term is
nonnegative. For a non-product state $\ket{\omega_{ABC}}$, its bipartite concurrence
$C_{A|BC}$ is a positive value and we have the CKW relation $C_{A|BC}^2\geq
C_{AB}^2+C_{AC}^2$. When $C_{AB}^2\neq0$ and $C_{AC}^2\neq 0$, we can obtain
that $T_1(E_f^2)>0$ because the entanglement $E_f^2(C^2)$ is a monotonically
increasing and convex function of the concurrence $C^2$. When one of the two-qubit
concurrence is zero, for example $C_{AC}^2=0$, the CKW relation is $C_{A|BC}^2>
C_{AB}^2$. According to the monotonic property, we have $T_1(E_f^2)>0$. It should
be noted that $C_{A|BC}^2=C_{AB}^2$ should be removed simply because it corresponds
to the case that the three-qubit pure state is a product one under the partition $AB|C$.
Therefore, $T_1(E_f^2)>0$ if ever the three-qubit state is of nonproduct, implying
that the measure $Q_3(A|BC)$ is positive.

So far, we have shown that the introduced tripartite quantum correlation measure
$Q_3(A|BC)$ satisfies all the three necessary criteria. Furthermore, the measure may
be understood as the monogamy score difference of SQD between the given state and a
bipartite product state, \emph{i.e.,}
\begin{eqnarray}\label{26}
    Q_3(A|BC)&=&||\psi_{ABC}-\varphi_{A}\otimes\varphi_{BC}||_{MD2}\nonumber\\
    &=& M_{D2}(\psi_{ABC})-M_{D2}(\varphi_{A}\otimes\varphi_{BC}),
\end{eqnarray}
where monogamy score is $M_{D2}(ABC)=D^2_{A|BC}-D^2_{A|B}-D^2_{A|C}$. When
$Q_3(A|BC)$ is nonzero, the quantum state is not a product state and its monogamy
score is larger than that of any bipartite product state. The score difference is
just the residual SQD. The larger the value of $Q_3(A|BC)$, the farther the
monogamy distance between the give state and the bipartite product state.
Therefore the measure $Q_3(A|BC)$ can characterize the genuine three-qubit quantum
correlation and has a physical explanation in terms of the monogamy score
difference.

\begin{figure}
\begin{center}
\epsfig{figure=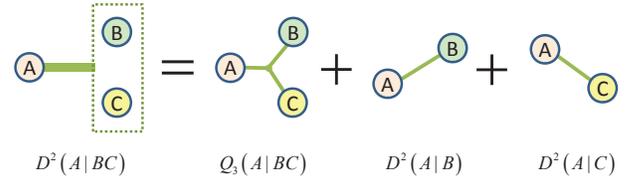,width=0.45\textwidth}
\end{center}
\caption{(Color online) The hierarchy structure of quantum correlations in a
three-qubit pure state.}
\end{figure}

In addition, for a three-qubit pure state $\ket{\psi_{ABC}}$, we can obtain a
hierarchy structure of quantum correlations. As depicted schematically in Fig.3,
Eq.(25) can be rewritten as
\begin{equation}\label{27}
D^2_{A|BC}=D^2_{A|B}+D^2_{A|C}+Q_3(A|BC),
\end{equation}
where $D^2_{A|BC}$ quantifies the total quantum correlation in the partition $A|BC$,
$D^2_{A|B}$ and $D^2_{A|C}$ quantify two-qubit quantum correlations, and $Q_3(A|BC)$
characterizes the genuine three-qubit quantum correlation under the partition
$A|BC$.

As an application, we consider generalized Greenberger-Horne-Zeilinger (GHZ) and $W$ states, which are
two inequivalent classes under stochastic local operations and classical
communication \cite{dur00pra}. The generalized $GHZ$ state has the form
$\ket{G_3}=\alpha\ket{000}+\beta\ket{111}$. Its two-qubit quantum correlations are
zero because the reduced density matrices $\rho_{ij}$ are classical states.
Therefore, there is only the genuine three-qubit quantum correlation $Q_3(A|BC)=S^2(A)$
in the generalized $GHZ$ state. For the generalized $W$ state
$\ket{W_3}=a\ket{001}+b\ket{010}+c\ket{100}$, both two-qubit and three-qubit
quantum correlations are nonzero when parameters $a$, $b$, and $c$ are nonzero. When
$a=b=1/2$ and $c=\sqrt{2}/2$, the tripartite quantum correlation has the maximal
value $Q_3(A|BC)\simeq 0.2779$.

Also noting that the QD is asymmetric for different measurement parties, the
tripartite quantum correlation under qubit permutation is not equivalent to each
other: $Q_3(A|BC)\neq Q_3(B|AC)\neq Q_3(C|AB)$ for a generic quantum state. From
this consideration, we may define a new tripartite quantum correlation measure:
\begin{equation}\label{28}
    Q_3(\ket{\psi_{ABC}})=\frac{1}{3}\sum_{i,j,k} Q_3(i|jk),
\end{equation}
where $i\neq j\neq k\in \{A,B,C\}$, and the measure may be referred to as the three-qubit
mean-SQD. This mean-SQD not
only satisfies all three conditions for a multipartite correlation measure, but also
is independent of bipartite partitions, reflecting really the global tripartite
quantum correlation in a three-qubit pure state $\ket{\psi_{ABC}}$.

\subsection{Tripartite correlation indicator in mixed states}

In three-qubit mixed states, the quantum correlation distribution of SQD is not
always monogamous. As an example, we analyze the quantum state
\begin{equation}\label{29}
    \rho_{ABC}(W)=\proj{\psi_1}+\proj{\psi_2},
\end{equation}
where the non-normalized pure state components are $\ket{\psi_1}=a\ket{100}+
b\ket{010}+c\ket{001}$ and $\ket{\psi_2}=d\ket{000}$, respectively. Using the
Koashi-Winter formula, we have the discord
\begin{equation}\label{30}
D_{A|BC}=E_f(AE)-S(A|BC)
\end{equation}
where subsystem $BC$ is equivalent to a logic qubit and the subsystem $E$ is the
environment degree of freedom purifying the mixed state. Because $\rho_{ABC}(W)$ is a
rank-2 quantum state, the environment subsystem is equivalent to a logic qubit. In
Eq. (29), we set the parameters $a=\mbox{cos}\theta_1$,
$b=\mbox{sin}\theta_1\mbox{sin}\theta_2\mbox{cos}\theta_3$,
$c=\mbox{sin}\theta_1\mbox{sin}\theta_2\mbox{sin}\theta_3$, and
$d=\mbox{sin}\theta_1\mbox{cos}\theta_2$. When the parameters
$\theta_1=\theta_2=\theta_3=0.4\pi$, we can get $E_f(AE)=0.06942$ by using the
Wootters formula \cite{wootters98prl}, which results in $D_{A|BC}^2=0.10845$.
Similarly, we have $D_{A|B}=0.02368$ and $D^2_{A|C}=0.08994$. Substituting these
SQDs into the correlation distribution $D^2_{A|BC}-D^2_{A|B}-D^2_{A|C}$, we can
determine the value of the distribution is $-0.00517$.

Although the quantum correlation distribution can be negative, we can still introduce
a tripartite quantum correlation indicator whenever the distribution
in a mixed state $\rho_{ABC}$ is always monogamous (an example of this case will
be presented in the next section). In this case, we may define the indicator as
\begin{equation}\label{31}
\mathcal{Q}_3(\rho_{i|jk})=D^2_{i|jk}-D^2_{i|j}-D^2_{i|k},
\end{equation}
where  $i\neq j\neq k\in \{A,B,C\}$. Furthermore, we can introduce a symmetric
tripartite correlation indicator
\begin{equation}\label{32}
\mathcal{Q}_3(\rho_{ABC})=\frac{1}{3}\sum_{i\neq j\neq k} \mathcal{Q}_3(i|jk),
\end{equation}
which indicates the global tripartite quantum correlation in a three-qubit mixed
state.

\section{Multipartite quantum correlation indicators in four-qubit systems}

In four-qubit pure states, the structure of quantum correlation distributions
is more complicated than that in three-qubit states. In general, these distributions
are not monogamous. However, if the distributions of SQD are monogamous in
a given four-qubit system, we can
also construct an indicator of the four-body correlation with the components
\begin{eqnarray}\label{33}
  && \mathcal{Q}_4^{(1*3)}=D^2_{A|BCD}-D^2_{A|B}-D^2_{A|C}-D^2_{A|D},\\
  && \mathcal{Q}_4^{(2*2)}=D^2_{AB|CD}-D^2_{A|C}-D^2_{A|D}-D^2_{B|C}-D^2_{B|D},\nonumber
\end{eqnarray}
where the superscript $(1*3)$ means that the correlation distribution lies in the
partition between one qubit and the other three qubits and the case for $(2*2)$ is
the distribution between two two-qubit subsystems.
Under qubit permutations, $\mathcal{Q}_4^{(1*3)}$ and $\mathcal{Q}_{4}^{(2*2)}$ have
four and six inequivalent components, respectively.
The nonzero component indicates the genuine multipartite quantum correlation
in the designated partition of a given state.
For example, in the generalized four-qubit $GHZ$ state $\ket{G_4}=\alpha\ket{0000}
+\beta\ket{1111}$, the correlation distribution is always non-negative, and we have
$\mathcal{Q}_4^{(1*3)}=\mathcal{Q}_4^{(2*2)}=S^2(A)$.
Another example is the cluster state
$\ket{C_4}=(\ket{0000}-\ket{0111}-\ket{1010}+\ket{1101})/2$
\cite{rau01prl}, in which we have $\mathcal{Q}_4^{(1*3)}=1$ and
$\mathcal{Q}_4^{(2*2)}=2$.

At this stage, as an interesting example, we consider the dynamical property of
quantum correlations in a real quantum system. As is known, the dynamical property
of a two-qubit quantum correlation has been widely investigated both theoretically and
experimentally (see, for example, Refs.
\cite{maz09pra,maz10prl,jxu10nc,auc11prl,cen12pra,rong12prb,fran13ijmpb} and
references therein). However, the dynamical property of multipartite quantum
correlations is still very challenging. We now use the multipartite correlation
indicator to analyze the dynamical evolution in four-partite cavity-reservoir
systems. The system is composed of two entangled cavity photons being affected by
the dissipation of two individual $N$-mode reservoirs, where the interaction of a
single cavity-reservoir system is described by Hamiltonian \cite{lop08prl}
\begin{equation}\label{34}
\hat{H}=\hbar \omega \hat{a}^{\dagger}\hat{a}+\hbar\sum_{k=1}^{N}\omega_{k}
\hat{b}_k^{\dagger}\hat{b}_k+\hbar\sum_{k=1}^{N}g_{k}(\hat{a}
\hat{b}_{k}^{\dagger}+\hat{b}_{k}\hat{a}^{\dagger}).
\end{equation}
The initial state is $\ket{\Phi_0}=(\alpha\ket{00}+\beta\ket{11})_{c_1c_2}
\ket{00}_{r_1r_2}$, where the dissipative reservoirs are in the vacuum state. In the limit of
$N\rightarrow \infty$ for a reservoir with a flat spectrum, the output state of the cavity-reservoir system has the form \cite{lop08prl}
\begin{equation}\label{35}
  \ket{\Phi_t}=\alpha\ket{0000}_{c_1r_1c_2r_2}+\beta\ket{\phi_t}_{c_1r_1}\ket{\phi_t}_{c_2r_2},
\end{equation}
where $\ket{\phi_t}=\xi(t)\ket{10}+\chi(t)\ket{01}$ with the amplitudes being
$\xi(t)=\mbox{exp}(-\kappa t/2)$ and $\chi(t)=[1-\mbox{exp}(-\kappa t)]^{1/2}$. For
the output state, we analyze its relevant components of the three- and four-partite
quantum correlation indicators $\mathcal{Q}_3$ and $\mathcal{Q}_4$ given in Eqs.
(31) and (33). Here, we use the method introduced by Chen \emph{et al.} for
calculating the quantum discord of two-qubit $X$ states (see the calculation in
the Appendix) \cite{che11pra}.

\begin{figure}
\begin{center}
\epsfig{figure=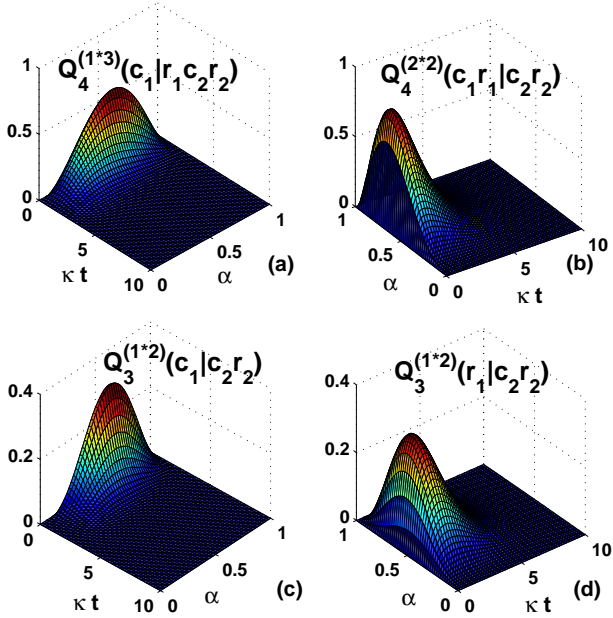,width=0.45\textwidth}
\end{center}
\caption{(Color online) Different components of multipartite quantum
correlation indicators in cavity-reservoir systems as a function of the
time evolution $\kappa t$ and the initial state amplitude $\alpha$,
where all the correlation distributions are non-negative and detect
the genuine multipartite quantum correlations.}
\end{figure}
\begin{figure}
\begin{center}
\epsfig{figure=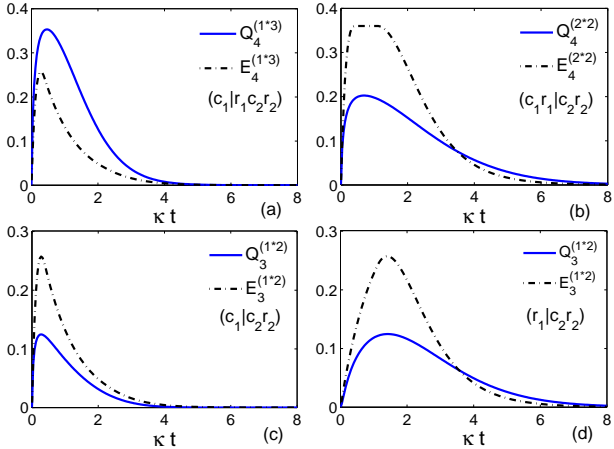,width=0.45\textwidth}
\end{center}
\caption{(Color online) The multipartite quantum correlation
indicators (blue solid lines) as a function of the time evolution parameter $\kappa t$
in comparison to the multipartite entanglement indicators (black dash-dotted lines) in the
output state $\ket{ \Phi_t}$ with the initial state parameter
$\alpha=1/\sqrt{10}$.}
\end{figure}

In Fig.4, we plot different components of multipartite quantum correlation indicators
as a function of
the time evolution parameter $\kappa t$ and the initial state amplitude $\alpha$. It
is noted that all the correlation distributions are non-negative and we have
$\mathcal{Q}_4\ge 0$ and $\mathcal{Q}_3\ge 0$ for these components. When the time
$\kappa t=0$, the quantum state is a product state and these indicators are zero.
Along with the time evolution, they first increase to their maxima, and then decay
asymptotically. When the parameter $\kappa t \rightarrow \infty$, the output state
evolves to a product state again and all the multipartite quantum correlations disappear.

In the cavity-reservoir system, its multipartite entanglement evolution
was investigated in Refs. \cite{lop08prl,byw09pra,wen11epjd}. The genuine
multipartite entanglement can be characterized by a series of entanglement
indicators. Here, in our analysis, we consider the following components:
\begin{eqnarray}\label{36}
  &&E_4^{(1*3)}(\ket{\Phi_t})=C_{c_1|r_1c_2r_2}^2-C_{c_1r_1}^2-C_{c_1c_2}^2-C_{c_1r_2}^2,
    \nonumber\\
  &&E_4^{(2*2)}(\ket{\Phi_t})=C_{c_1r_1|c_2r_2}^2-C_{c_1c_2}^2-C_{r_1r_2}^2-\sum
  C_{c_ir_j}^2,\nonumber\\
  &&E_3^{(1*2)}(\rho_{c_1c_2r_2})=C_{c_1|c_2r_2}^2-C_{c_1c_2}^2-C_{c_1r_2}^2,\nonumber\\
  &&E_3^{(1*2)}(\rho_{r_1c_2r_2})=C_{r_1|c_2r_2}^2-C_{r_1c_2}^2-C_{r_1r_2}^2,
\end{eqnarray}
where $C^2$ is the square of concurrence and the subscripts $i\neq j$ in the
second equation. The component $E_4^{(1,3)}$ can be used to characterize
the genuine multipartite entanglement in the partition $c_1|r_1c_2r_2$,
and $E_4^{(2,2)}$ can indicate the genuine block-block entanglement
in the partition $c_1r_1|c_2r_2$ \cite{byw09pra}. Moreover, the component
$E_3^{(1,2)}$ is used to quantify the qubit-block entanglement in
three-qubit mixed states \cite{loh06prl,byw08pra,wen11epjd}.

In Fig.5, we plot the relevant components of multipartite quantum correlation
indicators $\mathcal{Q}_4$ and $\mathcal{Q}_3$ in comparison to the
multipartite entanglement indicators $E_4$ and $E_3$ for the output state
$\ket{\Phi_t}$. As seen from the figure, the
multipartite quantum correlation is correlated with the
multipartite entanglement in every partition structure. However, the
peaks of correlation and entanglement do not
coincide completely. The reason is that quantum correlation and
quantum entanglement are not equivalent in general. Particularly, in
the dynamical procedure, the evolution of two-qubit entanglement can
exhibit the phenomenon of entanglement sudden death
\cite{hor01pra,eis03jmo,tyu04prl}, but the corresponding evolution
of quantum correlation is always asymptotic. In addition, the peak values of quantum
correlation indicators can be greater (Fig. 5a) or less (Fig. 5b-d) than those
of quantum entanglement indicators. This is due to the fact that different measures of quantum states
lack the same ordering \cite{virmani00pla,lang11ijqi,okrasa12epl}. Although
the quantum correlation can be greater than entanglement in separable states, the ordering
may change in a generic quantum state. For example, quantum discord is not always
greater than the entanglement of formation even in two-qubit quantum states \cite{luo08pra}.

\section{Discussion and conclusion}

The QD is very difficult to compute because of the minimization over
all positive operator-valued measures. Till now, the analytical
result of QD is still an open problem except for some specific
classes of quantum states
\cite{luo08pra,lang10prl,gio10prl,ade10prl,ali10pra,cen11pra,che11pra,shi12pra}.
However, in three-qubit pure states, we can calculate two-qubit QD via the
Wootters formula \cite{wootters98prl} and Koashi-Winter relation \cite{koashi04pra}.
In this case, the analytical formula of genuine tripartite quantum correlation
is available and can be rewritten as
\begin{eqnarray}\label{37}
    Q_3(A|BC)&=&S(A)^2-[E_f(AC)-S(A|B)]^2\nonumber\\
    &&-[E_f(AB)-S(A|C)]^2.
\end{eqnarray}
Therefore, in three-qubit pure states, not only the hierarchy structure of quantum
correlation holds but also all the quantum correlations can be calculated analytically.

In conclusion, we have explored multipartite quantum correlations
with the monogamy of SQD and answered the two important questions.
We have proven that the SQD is monogamous in three-qubit pure states and
the residual correlation is a reasonable measure for genuine three-qubit
quantum correlation, which gives a clear hierarchy structure for quantum
correlations. For three-qubit mixed states, although the distribution
of SQD is not always monogamous, we have constructed an effective indicator which can
detect the genuine tripartite quantum correlation in a specific class of states.
For four-qubit pure states, the monogamy property of SQD may still be
used to construct effective indicators for measuring genuine multipartite
quantum correlations. As an interesting example, we have addressed the evolution
of multipartite cavity-reservoir systems. The present work may shed a light on
understanding of quantum correlations in multipartite systems.

\section*{Acknowledgments}

This work was supported by the RGC of Hong Kong under Grant No. HKU7044/08P.
Y.K.B. and N.Z. were also supported by NSF-China (Grant No. 10905016), Hebei
NSF (Grant No. A2012205062), and the fund of Hebei Normal University. M.Y.Y.
was also supported by NSF-China (Grant No. 11004033) and  NSF of Fujian Province
(Grant No. 2010J01002).

\section*{Appendix: calculation of the discord in cavity-reservoir systems}

The density matrix of a two-qubit $X$ state can be written
\begin{equation}
  \rho^{AB}_X=\left(
              \begin{array}{cccc}
                a_{00} & 0 & 0 & a_{03} \\
                0 & a_{11} & a_{12} & 0 \\
                0 & a_{12}^* & a_{22} & 0 \\
                a_{03}^* & 0 & 0 & a_{33} \\
              \end{array}
            \right).
\end{equation}
When the elements satisfy the following relations \cite{che11pra}:
\begin{eqnarray}
  &&|a_{12}+a_{03}|\geq |a_{12}-a_{03}|,\nonumber\\
  &&|\sqrt{a_{00}a_{33}}-\sqrt{a_{11}a_{22}}|\leq |a_{12}|+|a_{03}|,
\end{eqnarray}
Chen \emph{et al} proved that the optimal measurement for the quantum discord is $\sigma_x$.
In the output state $\ket{\Phi_t}$, we find the optimal
measurement is $\sigma_x$ for state $\rho_{c_1c_2}$. Then, according to the definition
of the quantum discord in Eq. (4), we can get the value of $D_{c_1|c_2}^2$.
For other two-qubit quantum discords in the correlation distributions,
we find that the optimal measurement is also $\sigma_x$, where we use the property that
subsystem $c_ir_i$ ($i=1,2$) is equivalent to a logic qubit. In a similar way,
we can calculate these SQDs.

\end{document}